\documentclass[apj]{emulateapj}

\usepackage{graphicx}
\usepackage{apjfonts}
\usepackage{amsmath}
\usepackage[raggedright]{subfigure}
\usepackage{color}
\usepackage{hyperref}

\newcommand{\psr}{PSR~J2339$-$0533}
\newcommand{\zeroFGL}{0FGL~J2339.8$-$0530}

\newcommand{\Eref}[1]{Equation~(\ref{#1})}
\def\Fermi{\textit{Fermi}}

\newcommand{\Tasc}{T_{\rm ASC}}
\newcommand{\forb}{f_{\tiny{\rm orb}}}
\newcommand{\Porb}{P_{\rm orb}}
\newcommand{\phiorb}{\phi_{\rm orb}}
\newcommand{\Pmod}{P_{\rm mod}}
\newcommand{\pvec}{\textbf{u}}

\shorttitle{Gamma-ray Timing of \psr} 
\shortauthors{\sc Pletsch \& Clark}

\begin{document}

\title{
Gamma-ray timing of redback \psr{}: Hints for \\gravitational quadrupole moment changes
}

\author{
Holger~J.~Pletsch\altaffilmark{1,2,3} and Colin~J.~Clark\altaffilmark{1,2}
}
\altaffiltext{1}{Max-Planck-Institut f\"ur Gravitationsphysik (Albert-Einstein-Institut), D-30167 Hannover, Germany}
\altaffiltext{2}{Institut f\"ur Gravitationsphysik, Leibniz Universit\"at Hannover, D-30167 Hannover, Germany}
\altaffiltext{3}{email: holger.pletsch@aei.mpg.de}

\begin{abstract} 
\noindent
We present the results of precision gamma-ray timing measurements of the 
binary millisecond pulsar \psr{}, an irradiating system  of ``redback'' type,
using data from the \textit{Fermi} Large Area Telescope. 
We describe an optimized analysis method to determine a long-term phase-coherent timing solution
spanning more than six years, including a measured eccentricity of the binary orbit and constraints 
on the proper motion of the system.
A major result of this timing analysis is the discovery of an extreme 
variation of the nominal $4.6$\,hr orbital period $\Porb$ over time, 
showing alternating epochs of decrease and increase. 
We inferred a cyclic modulation of $\Porb$ with 
an approximate cycle duration of $4.2$ years and 
a modulation amplitude of $\Delta\Porb/\Porb = 2.3\times10^{-7}$. 
Considering different possible physical causes, 
the observed orbital-period modulation most likely results from  
a variable gravitational quadrupole moment of the companion star due to 
cyclic magnetic activity in its convective zone.
\end{abstract}


\section{Introduction}\label{s:intro}

The Large Area Telescope \citep[LAT;][]{generalfermilatref} 
on the \Fermi{} satellite has paved the way to greatly increase the known Galactic population
of compact binary systems harboring irradiating millisecond pulsars (MSP).
These intriguing systems are commonly referred to as ``black widows'' and ``redbacks'' 
\citep{Roberts2013}, as the intense pulsar radiation is gradually destroying the companion star. 
Typically, in black widows the mass of the degenerate companion is very low \mbox{($\sim0.008-0.05\,M_\odot$)},
whereas the redback type is distinguished by having a 
non-degenerate, main-sequence-like companion that is more massive \mbox{($\sim0.15-0.6\,M_\odot$)}. 
Both types of systems provide interesting opportunities to study
the interaction between the pulsar wind and the stellar companion,
their unusual formation history \citep{Chen+2013,Smedley+2014} and 
the evolutionary link to low-mass X-ray binaries (LMXBs) \citep{Archibald+2009,Papitto+2013},
as well as the masses of neutron stars \citep{Kerkwijk+2011,Romani+2012-J1311-2}.

Since the launch of \Fermi, numerous new black widow and redback pulsars were found  
in targeted pulsar searches of formerly unidentified LAT gamma-ray 
sources with radio telescopes \citep[e.g.,][]{Ransom2011,Ray+2012,Barr+2013}. 
Only in one case a direct search in LAT data revealed the gamma-ray pulsations of PSR J1311$-$3430 \citep{Pletsch+2012-J1311} by exploiting partial knowledge of the orbit obtained from prior identification of
the heated companion at optical and X-ray wavelengths \citep{Romani2012}. Although the pulsar was subsequently
detected in the radio band \citep{Ray+2013-J1311}, it was shown to be weak and intermittent,
which had precluded a detection in typical radio searches.

The discovery path to the redback system \psr{} followed a similar route. The formerly unidentified
LAT source \zeroFGL{} was one of the bright gamma-ray sources unveiled during
the first 3 months of the \Fermi{} mission \citep{FermiBSL}. X-ray and optical
follow-up observations identified a probable counterpart,
which showed many properties reminiscent of an MSP in a
$4.6$\,hr binary orbit around a low-mass companion  \citep{Romani+2011-J2339,Kong+2012}. 
Only pulsations were lacking for an unambiguous confirmation of the pulsar nature.
Using the Green Bank Telescope, \citet{Ray+2014-J2339,Ray+2015-J2339} detected 
$2.8$\,ms radio pulsations. They showed that the companion was substantially more 
massive than modeling of the optical data had initially suggested.
With the spin period and orbital parameters tightly constrained, they were also able
to detect gamma-ray pulsations with the LAT, verifying the identification of \zeroFGL{}.
Their observed dispersion measure (DM) from the radio detection provided
an estimated distance  of about $450$\,pc \citep{Ray+2014-J2339,Ray+2015-J2339}.
Up to now, a conjectured variability in the orbital parameters had precluded a phase-coherent ephemeris 
for the pulsar covering the entire time span of the available LAT data.

Other black widow and redback MSP systems have also been observed to show 
significant variations in their orbital parameters \citep{Arzoumanian+1994,Doroshenko+2001,Archibald+2013}. 
In the interest of better understanding such phenomena, it is essential to
carry out precision timing over longer time intervals \citep[e.g.,][]{2000ASPC..202...67N,Lazaridis+2011}. 
Often the eclipse of the pulsar's radio emission over a large fraction of
the orbit additionally complicates or even prohibits an accurate timing analysis
of the orbital-parameter variability.
On the contrary, the pulsar's gamma-ray emission is essentially 
unaffected by this problem.
Thus, the continuously recorded multi-year LAT survey-mode data is uniquely suited
to carefully monitor the long-term evolution of the orbital parameters of those systems.

Here, we present the results of a precision timing analysis of \psr{}
using LAT gamma-ray photon data that cover more than six years.
In Section~\ref{ss:analysis}, we describe the data preparation, 
the improved timing methodology, and provide the
details of the inferred phase-coherent pulsar ephemeris.
Section~\ref{s:discussion} provides a detailed discussion of these results.
Finally, a summarizing conclusion follows in Section~\ref{s:conclusions}.

\newpage

\section{Gamma-ray Data Analysis} \label{ss:analysis}

\subsection{Data Preparation}

For this analysis we employed Pass~7~reprocessed LAT photon data recorded between 
2008 August~4 and 2014 October~17. 
Via the \Fermi{} Science 
Tools\footnote{http://fermi.gsfc.nasa.gov/ssc/data/analysis/scitools/overview.html} 
we selected photons belonging to the \texttt{P7REP\_SOURCE} class
with reconstructed directions within~$8$\degr{} 
of the pulsar position, energies from $0.1$ to $100$\,GeV, and zenith angles $\leq 100$\degr.
Photons recorded when the LAT's rocking angle exceeded $52$\degr 
or when the LAT was not in nominal science mode were excluded.

We constructed spectral models including all gamma-ray sources found within $13$\degr{} of the 
nominal pulsar position from a preliminary version of the third Fermi LAT source catalog \citep[3FGL;][]{3FGL-Cat} 
based on four years of LAT data. 
The parameters were left free only for point sources within the inner $5$\degr{}.
The source models included contributions from the Galactic diffuse emission,
the extragalactic diffuse emission, 
and the residual instrumental background\footnote{http://fermi.gsfc.nasa.gov/ssc/data/access/lat/BackgroundModels.html}. 
The pulsar spectrum was modeled as an exponentially cutoff power law, 
$dN / dE \propto E^{-\Gamma} \exp\left( - E / E_c \right)$, 
where $\Gamma$ denotes the spectral index and $E_c$ is the cutoff energy. 
Using this spectral model of the region
and the \texttt{P7REP\_SOURCE\_V15} instrument response functions,
we employed \texttt{gtsrcprob} to compute a weight for each photon \citep{KerrWeightedH2011}, 
measuring the probability of having originated from the pulsar for further background suppression.

From the final data set we discarded photons with probability weights 
less than $0.001$.  
While this chosen weight cutoff 
makes only a negligible change to the significance of the detected pulsations, 
it dramatically improves the computational efficiency for the subsequent 
timing analysis.

\subsection{Timing Analysis}

A central aspect of this work is to obtain a precise ephemeris, 
i.e. a phase-coherent timing solution, covering the full time range of 
LAT data available. This means counting the exact number of pulsar 
turns over the past six years of the \Fermi{} mission.
The initial ephemeris obtained from the radio pulsar discovery 
using a circular-orbit model was sufficient to reveal significant gamma-ray 
pulsations \citep{Ray+2014-J2339}, but was unable to maintain accurate phase 
coherence over a time span longer than about three years, most likely
due to unaccounted orbital variability of the \psr{} system.
This motivated the present investigation to study and model the
putative orbital parameter variation, facilitating a long-term
phase-coherent timing solution.

To begin with, we examined the variation of the orbital period $\Porb$ and the
projected semimajor axis~$x$ across neighboring subsets of data.  
For this, we fixed the sky position to the location of the optical counterpart
at $\alpha = 23^{\rm h}39^{\rm m}38\fs75$ and $\delta = -05\arcdeg33\arcmin05\farcs3$
(J2000.0) from \citet{Romani+2011-J2339}.
The remaining pulsar parameters were set to those of the preliminary ephemeris. 
Assuming a circular orbit we scanned ranges in $\{\Porb, x\}$ on a dense grid
around their initial values at fixed time of ascending node $\Tasc$. At each grid point we computed  
the weighted $H$-test statistic \citep{deJaeger1989,KerrWeightedH2011} 
using photons within a fixed time window of size $110$~days.
This time window covered almost two precession periods of \Fermi{} ($56$~days) 
and was chosen by balancing signal-to-noise ratio and time resolution, 
being just long enough to still accumulate a detectable signal-to-noise ratio.
This window was then slid over the entire data set using a 50\%~overlap
between subsequent steps. 
 
The results of this study provided the first evidence that \psr{} undergoes alternating 
intervals of orbital-period increase and decrease,
whereas for $x$ no significant variation was observed.
To obtain an approximate solution for the evolution of $\Porb$
over time, we considered the $\Porb$ value giving the highest $H$-test 
in each sliding-window step. It became apparent that modeling 
this set of $\Porb$ values over time required us to employ a polynomial 
expansion of the orbital frequency $\forb=1/\Porb$ 
about $\Tasc$, so that for a given time~$t$ the orbital phase $\phiorb$ 
(measured in cycles) is written as
\begin{equation}
  \phiorb(t) = \sum_{k=0}^K \, \frac{\forb^{(k)}}{(k+1)!} \, \left( t-\Tasc \right)^{(k+1)} \,,
  \label{e:orbphase}
\end{equation}
where $\forb^{(k)}$ represents the $k$th time-derivative of $\forb$.
Simple least-square minimization suggested that the order $K$ needed to give
an adequate fit was at least five or six. Ultimately, to determine the model providing 
the best fit we used the more sensitive timing analysis described next.

\begin{deluxetable}{ll}[t]
\tablewidth{\columnwidth}
\tablecaption{\label{t:tim} Parameters for \psr }
\tablecolumns{2}
\tablehead{
\colhead{Parameter} &
\colhead{Value}
}
\startdata
Range of observational data (MJD) \dotfill  & $54683$ -- $56947$ \\[0.15em]
Reference epoch (MJD)\dotfill  & $55100.0$   \\[0.15em]
\cutinhead{Timing Parameters}
R.A., $\alpha$ (J2000.0)\dotfill & $23^{\rm h}39^{\rm m}38\fs74(1)$\\[0.15em] 
Decl., $\delta$ (J2000.0)\dotfill    & $-05\arcdeg33\arcmin05\farcs32(3)$\\[0.15em]
Proper motion in $\alpha$, $\mu_\alpha \cos\delta$ (mas yr$^{-1}$) \dotfill & $11(4)$ \\[0.15em]
Proper motion in $\delta$, $\mu_\delta$ (mas yr$^{-1}$) \dotfill & $-29(10)$ \\[0.5em]
Spin frequency, $f$ (Hz)\dotfill  & $346.71337922051(2)$\\[0.15em]			     
1st spin frequency derivative, $\dot f$ (Hz s$^{-1}$)\dotfill & $-1.6952(8)\times 10^{-15}$ \\[0.15em] 
2nd spin frequency derivative, $|\ddot f|$ (Hz s$^{-2}$)\dotfill & $< 2\times 10^{-26}$ \\[0.15em]      						      
Orbital frequency, $\forb$ ($10^{-5}$ Hz) \dotfill &$5.993873572(1)$ \\[0.15em]
1st derivative of orb. frequency, $\forb^{(1)}$ (Hz s$^{-1}$) \dotfill & $5.97(5) \times 10^{-19}$\\[0.15em]
2nd derivative of orb. frequency, $\forb^{(2)}$ (Hz s$^{-2}$) \dotfill & $1.77(4) \times 10^{-26}$\\[0.15em]
3rd derivative of orb. frequency, $\forb^{(3)}$ (Hz s$^{-3}$) \dotfill & $-1.06(2) \times 10^{-33}$\\[0.15em]
4th derivative of orb. frequency, $\forb^{(4)}$ (Hz s$^{-4}$) \dotfill & $-5.6(2) \times 10^{-41}$\\[0.15em]
5th derivative of orb. frequency, $\forb^{(5)}$ (Hz s$^{-5}$) \dotfill & $1.41(5) \times 10^{-48}$\\[0.15em]
6th derivative of orb. frequency, $\forb^{(6)}$ (Hz s$^{-6}$) \dotfill & $1.15(6) \times 10^{-55}$\\[0.15em]
Epoch of ascending node, $\Tasc$ (MJD) \dotfill & $55791.9182085(3)$\\[0.15em]
Projected semimajor axis, $x$ (lt-s) \dotfill &  $0.611656(4)$ \\[0.15em]
Derivative of proj. semimajor axis, $|\dot x|$ (lt-s s$^{-1}$) \dotfill & $ < 6\times10^{-14}$ \\[0.15em]
1st Laplace-Lagrange parameter, $\epsilon_1 $ \dotfill & $1(1)\times10^{-5}$ \\[0.15em]
2nd Laplace-Lagrange parameter, $\epsilon_2 $ \dotfill & $-21(1)\times10^{-5}$ \\[0.15em]
\cutinhead{Derived Parameters}
RMS timing residual ($\mu s$) \dotfill & $6.7$ \\[0.15em] 
Spin period, $P$ (ms) \dotfill & $2.8842267415473(2)$\\[0.15em]
1st spin period derivative, $\dot P$ (s s$^{-1}$) \dotfill & $1.4102(6)\times 10^{-20}$\\[0.15em]
2nd spin period derivative, $|\ddot P|$ (s$^{-1}$) \dotfill & $<2\times 10^{-31}$\\[0.5em]
Orbital period, $\Porb$ (d) \dotfill & $0.1930984018(3)$\\[0.15em]	
1st derivative of orbital period, $\dot{P}_{\rm orb}$ (s s$^{-1}$) \dotfill &$-1.66(1)\times 10^{-10}$ \\[0.15em]	
Eccentricity, $e$ \dotfill & $2.1(1) \times 10^{-4}$\\[0.15em] 
\cutinhead{Gamma-ray Spectral Parameters}
Photon index, $\Gamma$ \dotfill & $1.2 \pm 0.3$ \\[0.15em]
Cutoff energy, $E_c$ (GeV) \dotfill  & $3.7 \pm 1.4 $  \\[0.15em]
Photon flux\tablenotemark{a} $F_{100}$ ($10^{-8}$ photons cm$^{-2}$ s$^{-1}$)\dotfill & $2.0 \pm 0.2$ \\[0.15em]
Energy flux\tablenotemark{a} $G_{100}$ ($10^{-11}$ erg cm$^{-2}$ s$^{-1}$) \dotfill & $2.8 \pm 0.1$ \\[-0.15em]
\enddata
\tablecomments{
Numbers in parentheses are statistical $1\sigma$ uncertainties in the last digits.
The JPL DE405 solar system ephemeris has been used and times refer to~TDB.}
\tablenotetext{a}{Measured over the energy range from $100$\,MeV to $100$\,GeV. }
\end{deluxetable}

\begin{figure*}
\centerline{
\hfill
\includegraphics[width=0.99\textwidth]{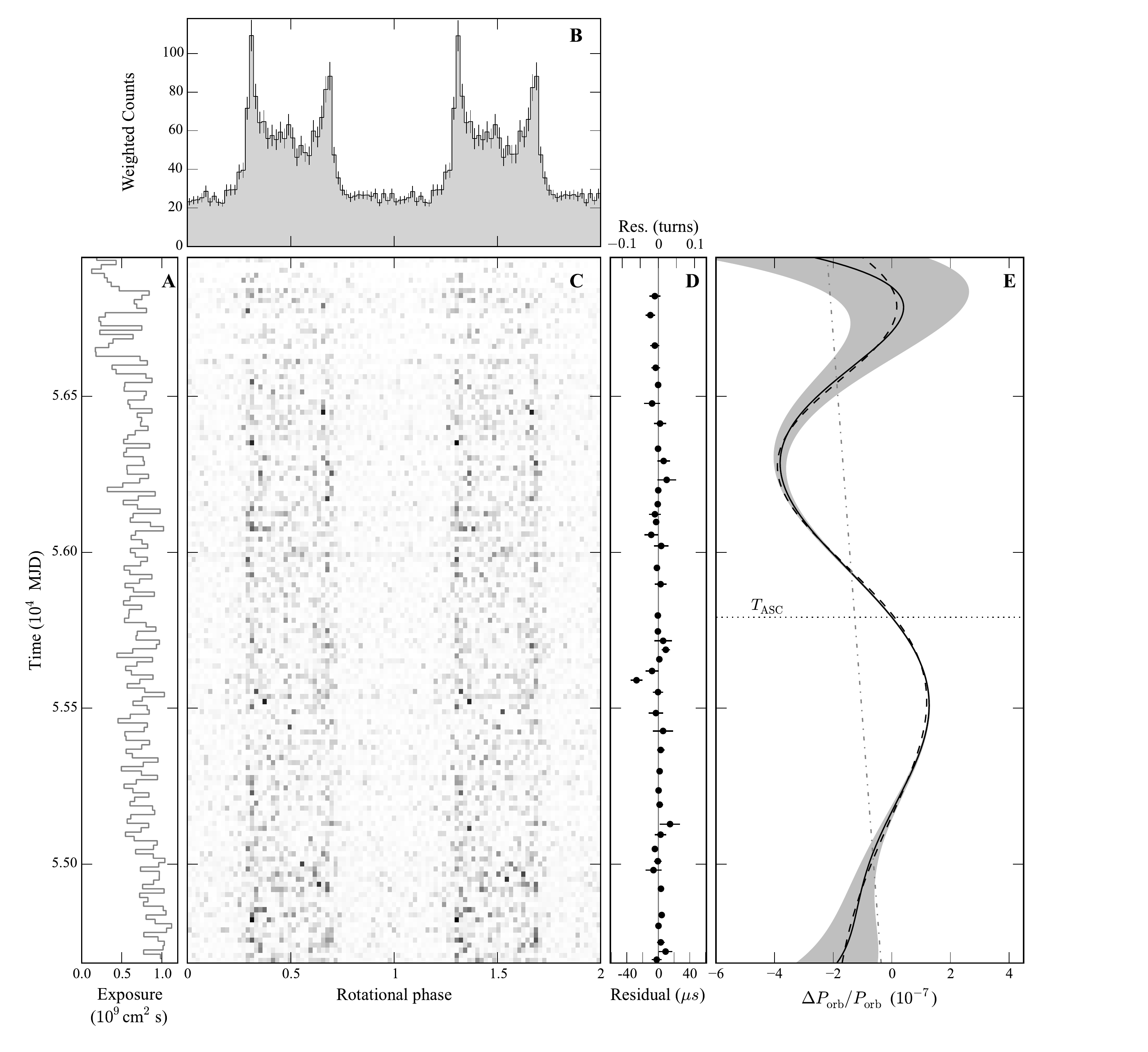}
\hfill
}
\caption{\label{f:tim} 
Panel~A: exposure versus time, binned in steps of 14 days.
Panel~B: integrated weighted pulse profile, using 50 bins per rotation. 
The vertical error bars show statistical $1\sigma$ uncertainties. 
Two rotational cycles are shown for clarity.
Panel~C: two-dimensional weighted histogram of rotational phase versus time, 
using 140 bins over time. 
The weighted counts are not exposure-corrected.
Panel~D: formal timing residuals, where the error bars show statistical $1\sigma$ uncertainties. 
The data points were obtained from non-overlapping subsegments of data that gave approximately the
same signal-to-noise ratio (i.e. an $H$-test value of about 15).
Only significant measurements with a log-likelihood greater than $8$ were included.
Panel~E: time evolution of the orbital-period change (solid black curve) based on the  
timing solution of Table~\ref{t:tim} and the surrounding gray-shaded region shows the statistical $1\sigma$
uncertainties. The uncertainties increase with the time distance to $\Tasc$ (shown by the horizontal dotted line), 
since the polynomial expansion of the orbital-frequency used in the timing model is about $\Tasc$. 
The dashed curve represents the best-fit cyclic modulation model discussed 
in Section~\ref{ss:GQC} and the dotted-dashed
line shows the linearly changing component of this model. 
}
\end{figure*}

As a starting point for the precision timing analysis,
we extended the initial pulsar ephemeris by including six orbital-frequency
derivatives based on the approximation for the orbital-period 
evolution of \psr{} obtained in the previous step.  Since the
orbit is apparently of extremely low eccentricity~$e$, we used the Lagrange-Laplace
parametrization \citep{Lange+2001} to describe the binary motion with
$\epsilon_1 \equiv e \sin \omega$ and $\epsilon_2 \equiv e \cos \omega$, such that
$e  = (\epsilon_1^2 + \epsilon_2^2)^{1/2}$.
The construction of a pulse profile template followed the method of \citet{Fermi2PC}
extending the work of \citet{Ray+2011} by also including the photon weights,
labeled $w_j$ for the $j$th photon. 
Since this requires a valid ephemeris,
we first used only the subinterval (about half) of the full data set over 
which the initial ephemeris maintained phase coherence for this purpose. 
In general the pulse profile 
$F(\phi)$ can be represented by a wrapped probability density function (PDF) 
of the pulsar's rotational phase \mbox{$\phi\in[0,1)$}, measured in turns.
For \psr{}, we found that this PDF is well approximated by a sum of a constant 
background and five unimodal wrapped Gaussian distributions \citep[cf.][]{Fermi2PC}. 
The rotational phase~$\phi(t_j,\pvec)$ is a function of the photon arrival time~$t_j$
and the vector~$\pvec$ that collects all pulsar spin, positional, and orbital parameters.
Thus, to obtain the best-fit values for~$\pvec$ of a given timing model, 
we evaluated the log-likelihood \citep{Fermi2PC},
\begin{equation}
  \log \mathcal{L} (\pvec) = \sum_{j=1}^N \log\left[  w_j \, F(\phi_j(t_j,\pvec)) + (1-w_j) \right]\,,
  \label{e:loglike}
\end{equation}
where $N$ is the total number of gamma-ray photons.

To find the parameters that give the global maximum of the log-likelihood,
we followed a novel approach that differs from the method of \citet{Ray+2011}.
Note that in \citet{Ray+2011} the LAT data were subdivided into segments in time and
over each such segment the photon times were folded using a preliminary ephemeris
to obtain a pulse time-of-arrival (TOA) measurement. The set of TOAs from all segments
then served as input to a global fitting procedure. Due to the sparseness of the gamma-ray
photons, a sufficiently high signal-to-noise ratio for a single TOA determination
required long integration times of order weeks. However, this makes it very
difficult to detect effects on much shorter time scales, and orbital parameter variability in particular,
since the preliminary ephemeris used to fold the photon times 
in each segment is not necessarily accurate.
Given the pulsed flux level of \psr{}, one would need segments of duration at least 30 days. 
For comparison, this interval spans already about 155 orbital revolutions.

Therefore, we instead evaluated the log-likelihood of \Eref{e:loglike} \emph{directly} 
using the entire set of unfolded gamma-ray photon times (without binning in time). 
For the exploration of $\log \mathcal{L}$ over the relevant parameter space
of the timing model 
we employed the Monte Carlo (MC) sampler {\sc MultiNest} \citep{Feroz+2009-Multinest}.
This multimodal nested sampling algorithm is especially efficient in sampling 
challenging likelihood surfaces and allows
one to calculate posterior distributions as a byproduct.

The timing analysis remains an iterative procedure, specifically since the
actual pulse profile is not known a priori.  
In the first MC run we still held the sky position fixed to the optical-counterpart 
location \citep{Romani+2011-J2339}, but let
the spin and orbital parameters vary. This gave a first phase-connected solution
spanning the full LAT data set, from which we then refined the pulse profile template.
In the subsequent MC run, we then also let the sky location vary
and determined the best-fit values for $\pvec$, 
from which a further refined pulse profile template was derived for use in the next MC run.
Subsequently, we further extended the timing solution to also include
additional effects, such as proper motion and orbital eccentricity. 
We also tested for a variation in the projected semimajor axis, $\dot x$, and a
second spin-frequency derivative, $\ddot f$, but in both
cases no significant measurement was made and so we can only give upper limits.

Table~\ref{t:tim} presents the final timing solution, 
obtained after several iterations of pulsar-model and pulse-profile-template refinements. 
For the best estimate and uncertainty of each parameter in the timing model 
we report the mean value and standard deviation of the one-dimensional marginalized 
posterior distribution, as determined by the {\sc MultiNest} algorithm.
This timing model includes the polynomial expansion of the oscillating orbital frequency
as given in \Eref{e:orbphase} truncated at the sixth order ($K=6$), which gave a maximum
$\log \mathcal{L}$ value of about 580. 

We also repeated the timing procedure for different models with polynomial expansions truncated 
at orders $K=5$ and $K=7$. To determine which of the three models better fits the data, we then 
invoked the Bayesian Information Criterion \citep[BIC; ][]{Schwarz1978} as a comparative tool.
The BIC value is based on the number of free parameters to be estimated, the number
of data points and the maximized value of $\log \mathcal{L}$. The smaller BIC value is preferred, 
since the BIC becomes increasingly large when either the data is poorly fit or when the number of free 
parameters increases and the data is overfit. Thus, the BIC is minimized for the 
simplest model that sufficiently fits the data. In our case, the $K=7$ model gave about the same
$\log \mathcal{L}$ value as for $K=6$, but given the extra parameter, the simpler $K=6$
model is preferred according to the BIC. On the other hand, the $K=5$ model has one parameter
less compared to $K=6$, but gave a significantly lower $\log \mathcal{L}$ value,
so that the BIC again preferred the $K=6$ model. Therefore, we considered the model 
with $K=6$ as shown in Table~\ref{t:tim} as the optimal representation of the data.
However, we also caution that outside the time span of this data, the derived timing solution, 
specifically the polynomial expansion for the orbital frequency, likely has 
little predictive power.

\subsection{Timing Results}

In Figure~\ref{f:tim}, the inferred parameters of the final timing solution of Table~\ref{t:tim}
were used to generate the integrated pulse profile (panel~B), 
the diagram of rotational phase versus time (panel~C), 
and the residuals over time (panel~D). 
The rectilinearity of the phase tracks over time
and the smallness of the residuals clearly certify the phase-coherence of
the timing solution, accounting for the correct number of pulsar turns during
the observational time span.
Merely to illustrate the quality of the timing solution, in Figure~\ref{f:tim} (panel~D) we also show 
the phase residuals that have an rms about $1.5$\% in turn,
which translates into an rms timing accuracy of about $7\,\mu\textrm{s}$.

While the sky position of the final timing solution 
is compatible with the optical counterpart location, we were also able to
measure significant values for the proper motion of the system relative to the solar system 
barycenter ($\mu_\alpha,\mu_\delta$, given in Table~\ref{t:tim}). 
This amounts to a total transverse proper motion of 
$\mu_t = (\mu_{\alpha}^{2} \cos^2\delta + \mu_{\delta}^{2})^{1/2} \approx (31\pm 10)$\,mas yr$^{-1}$.
Combined with the radio-DM distance estimate of $d=0.45$\,kpc, 
we derive a transverse velocity of \mbox{$\textsl{v}_t = d\,\mu_t \approx30\,\textrm{km}\,\textrm{s}^{-1}$}.
This transverse movement contributes  
to the observed spin parameters $P$ and $\dot P$
due to a changing Doppler shift \citep{Shklovskii1970}
via: \mbox{$\dot P = \dot{P}_{\rm int} + P\textsl{v}_t^2/(d c)$}. 
The latter term, representing the Shklovskii effect,
 is about $22$\% of the observed~$\dot P$, so that the intrinsic spin-period derivative is
estimated as $\dot{P}_{\rm int}  \approx  1.1\times10^{-20}$\,s s$^{-1}$.
Hence, we derive a Shklovskii-corrected pulsar spin-down luminosity of
\mbox{$\dot E =4\pi^2I\,\dot{P}_{\rm int}/P^3 \approx 1.8\times10^{34}$\,erg s$^{-1}$},
where $I=10^{45}$\,g\,cm$^{2}$ is an assumed fiducial neutron star moment 
of inertia.

The timing solution also provides constraints
on the companion mass~$M_{\rm c}$ through the pulsar mass function by 
combining the measurements of $x$ and $\Porb$ as
\begin{align}
 f(M_{\rm c},M_{\rm p}) &= \frac{M^3_{\rm c}\; \sin^3 \iota}{(M_{\rm c}+M_{\rm p})^2} = \frac{4\pi^2\,x^3}{G\;\Porb^2}\nonumber\\ 
  &= (6.58942\pm 0.00012)\times 10^{-3}\,M_\odot\,,
\end{align} 
where $G$ denotes the gravitational constant, $M_{\rm p}$ labels the pulsar mass,
and $\iota$ is the inclination angle.  
When further combined with the optical observations of the companion 
mass estimates for both components are possible,
as will be elucidated in Section~\ref{ss:masses}.

We also measured an orbital eccentricity of \mbox{$e = 2.1(1)\times10^{-4}$}.
Given the short $4.6$\,hr orbital period of \psr{}, this value of residual eccentricity
is higher than predicted by the convective-fluctuation theory of \citet{Phinney1992}.
This may indicate additional evidence for convective fluctuations in the companion,
as we discuss in Section~\ref{ss:GQC}. While these are perhaps more complicated 
than assumed by \citet{Phinney1992}, they could have 
inhibited perfect circularization of the binary orbit of \psr{}.

Most remarkably, the timing solution unveils the extreme orbital-period 
variations of \psr{} over the six years of LAT data. 
The two most extreme values of the orbital-period derivative 
attained during this interval are \mbox{$\dot{P}_{\rm orb} = -5.8\times10^{-10}$\,s s$^{-1}$} 
and $\dot{P}_{\rm orb} = 2.7\times10^{-10}$\,s s$^{-1}$.
Panel~E in Figure~\ref{f:tim} displays the dramatic evolution of the fractional 
changes $\Delta\Porb/\Porb$ based on the ephemeris of Table~\ref{t:tim}. 
A  significant cyclic modulation of the orbital period is revealed, where
the observational interval appears to cover about one and a half cycles.
The possible causes of the $\Porb$-modulation are discussed 
in detail in Section~\ref{s:discussion} below.

\section{Discussion}\label{s:discussion}

\subsection{Component Masses}\label{ss:masses}

\begin{figure}[t]
\centerline{
\hfill
\includegraphics[width=0.98\linewidth]{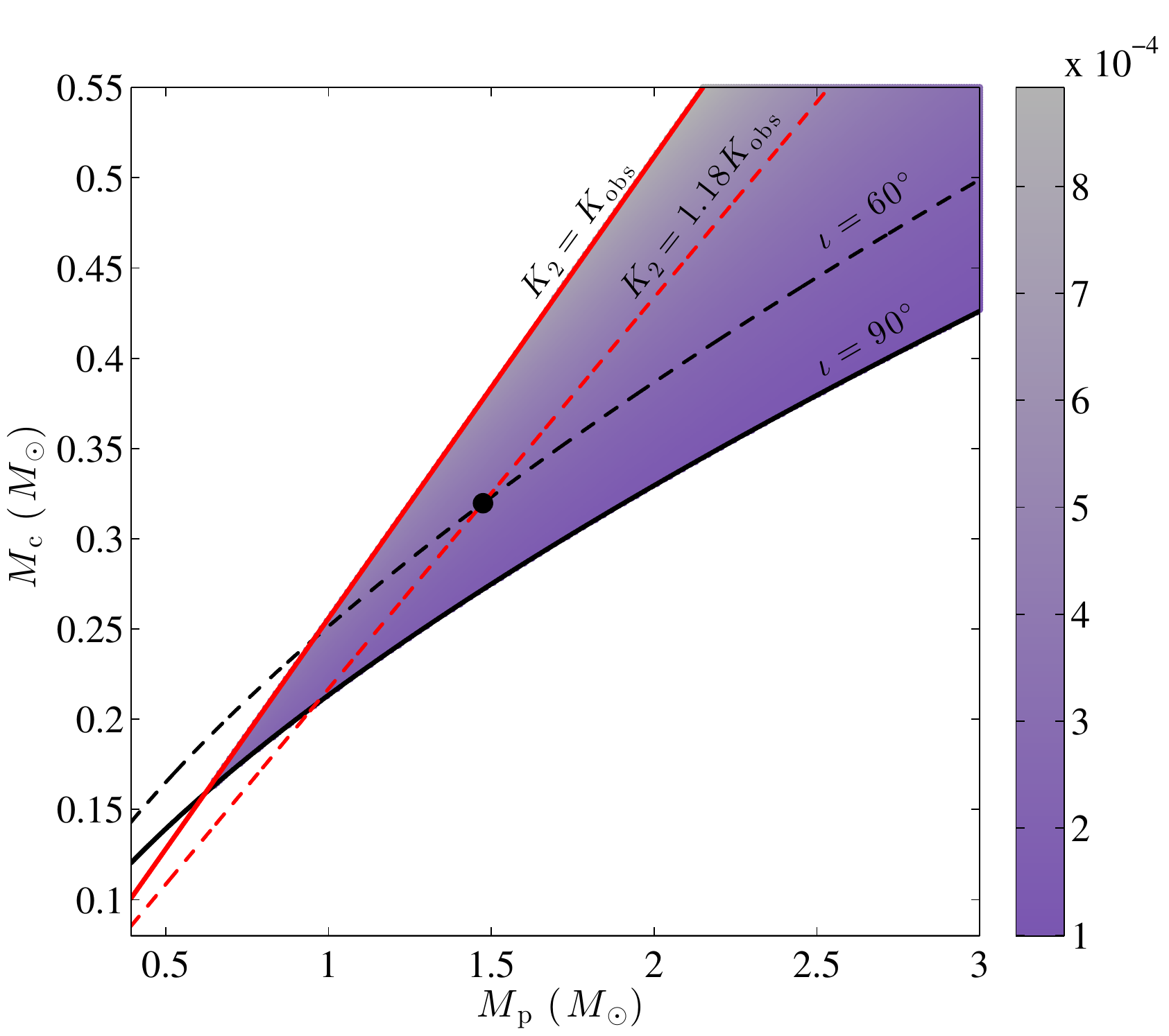}
\hfill
}
\caption{\label{f:massmass} 
Mass-mass diagram for the \psr{} binary system.
The non-shaded region is excluded by the pulsar mass function from the
gamma-ray timing solution since $\sin\iota\leq1$ (lower exclusion region)
and from the companion mass function based on 
the optical radial-velocity measurements \citep{Romani+2011-J2339} 
since $K_2\geq K_{\rm obs}$ (upper exclusion region).
Over the shaded (permitted) region, the color code shows how the fraction of the companion's
luminosity $\Delta L/L$ of \Eref{e:DeltaL} required
by the gravitational quadrupole coupling model varies with the assumed
component masses. The black dot indicates the fiducial values for the component masses 
assumed here, based on the optical lightcurve modeling by \citet{Romani+2011-J2339}.
\vspace*{0.25cm}
}
\end{figure}

By combining the optical radial velocity measurements of the irradiated 
companion \citep{Romani+2011-J2339} and the pulsar ephemeris,
further system parameters can be constrained, such as
the mass of the pulsar~$M_{\rm p}$ and of the companion~$M_{\rm c}$.
In forthcoming work that will exploit the extremely accurate pulsar
mass function when modeling the optical lightcurve,
more precise limits on the system parameters will be possible. 
However, here we employ the previous estimates by \citet{Romani+2011-J2339} 
in order to obtain fiducial values that are sufficient
for the purpose of the subsequent discussion and 
conclusion as will be shown below.

From radial velocity measurements of the companion, 
\citet{Romani+2011-J2339} observed a semi-amplitude 
of $K_{\rm obs}\approx270$\,km\,s$^{-1}$.
While these measurements track the apparent center of light, 
the radial-velocity amplitude of the companion's center of mass, 
$K_2$, can be larger \citep[e.g.,][]{Kerkwijk+2011}. 
The "$K$-correction" for this effect inferred by \citet{Romani+2011-J2339} 
gave $K_2 \approx 1.18\, K_{\rm obs}\approx320$\,km\,s$^{-1}$, which we adopt here.
The amplitude of the optical modulation mainly depends on the
inclination angle, which \citet{Romani+2011-J2339} estimated  
as $\iota\approx60$\degr. 
As shown in Figure~\ref{f:massmass}, 
this gives rise to the following estimates: $M_{\rm p} = 1.48\,\textrm{M}_\odot$ 
and $M_{\rm c} = 0.32\,M_\odot$, that we assume as fiducial values for the rest 
of this paper. The resulting mass ratio would be $q=M_{\rm p}/M_{\rm c}=4.61$.

For these mass estimates, the orbital separation is \mbox{$a \approx 1.71\,R_\odot$} and the
companion Roche lobe radius according to \citet{Eggleton1983} is:
$R_L \approx 0.88\,R_\odot$.
The actual radius of the companion star, $R_{\rm c}$, is difficult to obtain,
but can also be estimated from the optical light-curve modeling. 
In doing so, \citet{Romani+2011-J2339} found that 
the companion is close to Roche lobe filling, with a filling factor 
of $R_{\rm c}/R_{L} \approx 0.9$.
Further support for this picture is provided by the observed radio eclipses during
a large fraction of the orbit \citep{Ray+2014-J2339,Ray+2015-J2339}.
Thus, this implies a companion radius of $R_{\rm c} \approx 0.79\,R_\odot$,
which we also adopt as fiducial value for the remainder of this discussion.

The optical spectroscopy by \citet{Romani+2011-J2339} also shows that 
the companion's spectrum resembles that of a late-type star. 
However, the above estimates suggest that the stellar companion is likely less dense 
than a main-sequence star of the same mass,
which has also been found for other redback systems \citep[][]{Archibald+2013,Li+2014}.

\subsection{Energetics and Irradiation}\label{ss:energetics}

From the measured LAT energy flux $G_{100}$ (given in Table~\ref{t:tim}), 
we can obtain an estimate of the gamma-ray luminosity 
$L_\gamma=4\pi\,d^2\,G_{100} \approx 6.7\times10^{32}$\,erg s$^{-1}$, 
assuming no geometric beaming correction.
This leads to a gamma-ray conversion efficiency of 
$\eta_\gamma = L_\gamma/\dot E \approx 4$\%,
not atypical compared to the MSPs studied in~\citet{Fermi2PC}.

We can also estimate the effect of pulsar irradiation on the companion.
\citet{Romani+2011-J2339} inferred a companion-backside temperature 
of approximately $2800$\,K, which suggests an
intrinsic luminosity of $L \approx 1.3\times10^{32}$\,erg\,s$^{-1}$ ($0.034\,L_\odot$)
based on the above estimated radius. 
The irradiation efficiency $\eta_{\rm irr}$ is often used to describe the
fraction of the pulsar spin-down energy absorbed and converted into optical emission.
Assuming the contribution of an isotropic pulsar wind 
to the companion's optical luminosity via re-radiation
is \mbox{$L_{\rm irr} = \eta_{\rm irr} \dot E R_{\rm c}^2/(4a^2) \approx  9.7 \times10^{32} \eta_{\rm irr}$\,erg\,s$^{-1}$},
we find an  efficiency of $\eta_{\rm irr} = L/L_{\rm irr} \approx 14$\%.
This is within the range of typical irradiation efficiencies as found for the 
systems investigated in \citet{Breton+2013}.

\subsection{Possible Causes for the Orbital Period Variations}\label{ss:Porbmod}

Generally, an observed rate of change in the binary orbital period, $\dot{P}_{\rm orb}$, may  
result from various physical effects as discussed in \citep{Doroshenko+2001,Lazaridis+2011}.
The possible causes, most relevant regarding the time scale of our observations,
can be summarized as
\begin{equation}
  \dot{P}_{\rm orb} = \dot{P}_{\rm orb}^{\rm GW} 
  + \dot{P}_{\rm orb}^{\rm D} + \dot{P}_{\rm orb}^{M} 
  + \dot{P}_{\rm orb}^{\rm Q}  \,,
  \label{e:Porbdot}
\end{equation}
including energy loss due to gravitational-wave emission ($\dot{P}_{\rm orb}^{\rm GW}$),
changes due to Doppler shifts ($\dot{P}_{\rm orb}^{\rm D}$),
mass loss from the system ($\dot{P}_{\rm orb}^{M}$),
and gravitational quadrupole moment coupling ($\dot{P}_{\rm orb}^{\rm Q}$).

The term $\dot{P}_{\rm orb}^{\rm GW}$ in \Eref{e:Porbdot} is the 
contribution due to gravitational-wave emission. 
For the case of circular orbits this is given by \citep{Peters1964},
\begin{equation}
  \dot{P}_{\rm orb}^{\rm GW} = -\frac{192\pi}{5} 
     \left[ \frac{2\pi}{P_{\rm orb}}\,\frac{G M_{\rm c}}{c^3} \right]^{5/3}
    \frac{q}{(q+1)^{1/3}} \,,
    \label{e:PorbGW}
\end{equation}
where $c$ denotes the speed of light.
The resulting value for $\dot{P}_{\rm orb}^{\rm GW}\simeq -1.3\times10^{-13}$\,s s$^{-1}$  
is about three orders of magnitude smaller than 
the measured values of $\dot{P}_{\rm orb}$, 
and hence unlikely to be the primary cause.

The term $\dot{P}_{\rm orb}^{\rm D}$ in \Eref{e:Porbdot} represents the combined effect of 
proper motion \citep{Shklovskii1970} and Doppler shifts due to a changing distance to the binary
system, e.g., from the Galactic acceleration or due to a massive third body:  
\begin{equation}
  \dot{P}_{\rm orb}^{\rm D} = \dot{P}_{\rm  orb}^{\rm Shk} + \dot{P}_{\rm orb}^{\rm Gal} + \dot{P}_{\rm orb}^{\rm acc}\,.
\end{equation}
Using the measured proper-motion values given in Table~\ref{t:tim} 
and the radio-based DM distance of $d=0.45$\,kpc,
one obtains
\begin{equation}
  \dot{P}_{\rm orb}^{\rm Shk} = \frac{(\mu_{\alpha}^{2} \cos^2\delta +   \mu_{\delta}^{2})\, d}{c}\,P_{\rm orb} 
  \simeq 1.8\times10^{-14}\,\textrm{s}\,\textrm{s}^{-1} \,.
  \label{e:PorbShk}
\end{equation}
Contributions from the Galactic acceleration, $\dot{P}_{\rm orb}^{\rm Gal}$, 
are typically of similar magnitude \citep{Lazaridis+2009}.
An acceleration of the binary system caused by 
a massive third body is also highly unlikely. If that was the case, the pulsar spin
period and period derivative would be affected in the same way. From the measured
values in Table~\ref{t:tim} we can estimate the maximum contribution of this effect 
by assuming the apparent spin-down is entirely due to acceleration:
$(\dot{P}_{\rm orb}^{\rm acc}/\Porb) = (\dot P/P) \simeq 5\times10^{-18}$\,s$^{-1}$, implying 
\mbox{$\dot{P}_{\rm orb}^{\rm acc}\simeq10^{-14}$\,s s$^{-1}$}.
Thus, overall the term $\dot{P}_{\rm orb}^{\rm D}$ is found to be more than four orders of magnitude 
smaller than the observed $\dot{P}_{\rm orb}$ value for~\psr{}. 

The term $\dot{P}_{\rm orb}^{M}$ in \Eref{e:Porbdot} 
results from mass loss of the binary system. 
The spin-down luminosity of the pulsar irradiating the companion can drive
mass loss at a rate $\dot M_{\rm c}$ from the companion through an evaporative wind.
Taking the orbit to be circular and assuming no mass is lost from the pulsar,
one obtains \citep{Jeans1924},
\begin{equation}
  \dot{P}_{\rm orb}^{M} = -2 \frac{\dot{M}_{\rm c}}{M} \Porb\,,
  \label{e:PorbM}
\end{equation}
where $M = M_{\rm c} + M_{\rm p}$ denotes the total mass of the system.
Assuming that all of the radiative energy from the pulsar
intercepted by the companion (for a presumed isotropic pulsar wind) 
is converted into mass loss from the system,
this would imply a rate \mbox{$\dot{M}_{\rm c}\simeq 2.0\times10^{-8}\,M_\odot$yr$^{-1}$}
\citep{Stevens+1992}.
Inserting this value into \Eref{e:PorbM} gives \mbox{$\dot{P}_{\rm orb}^{M} \simeq -1.2\times10^{-11}$\,s s$^{-1}$}, 
which is more than an order of magnitude smaller in number than the measured $\dot{P}_{\rm orb}$.
More realistically, adopting the $14$\% efficiency for this process as estimated in Section~\ref{ss:energetics},
one gets   \mbox{$\dot{M}_{\rm c}\simeq 2.8\times10^{-9}\,M_\odot$yr$^{-1}$}
and \mbox{$\dot{P}_{\rm orb}^{M} \simeq -1.7\times10^{-12}$\,s s$^{-1}$}. This
therefore appears unlikely to be the primary cause for the orbital variation observed.

The possible contributions to $\dot{P}_{\rm orb}$ considered so far seem unlikely 
to be able to account for the large \mbox{$\Porb$-variation} measured.
In addition, these processes generate monotonic changes, which cannot
explain the alternating increase and decrease of $\Porb$ observed.
Hence this must be caused by the last term, $\dot{P}_{\rm orb}^{\rm Q}$, resulting 
from cyclic changes of the companion's gravitational quadrupole moment. 
In the following, we scrutinize the plausibility of this explanation,
confronting a specific theoretical model with the observational data.

\subsection{Gravitational Quadrupole Coupling}\label{ss:GQC}
 
To explain the observed orbital-period modulation, 
the only plausible cause of those discussed above that remains is a changing
gravitational quadrupole moment of the companion star
\citep{Matese+1983}. Such gravitational quadrupole coupling (GQC)  
likely results from magnetic activity in the stellar companion \citep{Applegate+1987,Applegate1992}. 
\citet{Applegate+1994} successfully applied the model by \citet{Applegate1992} 
to the black widow pulsar binary PSR~B1957+20,
gravitationally coupling the orbital-period changes reported in \citep{Arzoumanian+1994}
to changes in the quadrupole moment of the companion star.  
Similar $\Porb$ variations on a time scale comparable to that found for \psr{}
were seen in another black widow system, PSR~J2051$-$0827, 
and were also attributed to GQC \citep{Doroshenko+2001,Lazaridis+2011}.
The transitional redback system PSR~J1023$+$0038, which recently changed its state 
from MSP back to LMXB \citep{Stappers+2014}, also displayed
orbital-period changes at a comparable level \citep{Archibald+2013}. Further fitting into this picture,
notably analog orbital-period changes were found in long-term monitoring of LMXB systems  \citep{Wolff+2009,Patruno+2012}, likely also caused by GQC arising from cyclic magnetic activity.

In Applegate's GQC model  
the companion star is magnetically active with a convective envelope
and the pulsar is treated as a point mass moving in the gravitational field of the
companion in a circular orbit. When the companion's gravitational quadrupole moment
(due to tidal and centrifugal deformations) varies with time, this can cause
a variable orbital motion at constant orbital angular momentum. 
When the companion becomes more oblate, its
quadrupole moment increases, to balance gravitation the centripetal acceleration 
on the pulsar must increase, so that $\Porb$ must decrease, and vice versa.
Therefore, in principle any mechanism that can modulate the quadrupole moment 
of the companion also modulates the binary orbital period.

The specific mechanism suggested by \citet{Applegate1992} considers
cyclic, solar-like magnetic activity in the convective zone of the non-degenerate 
companion star. 
The magnetic field within the companion is assumed to generate a torque, which
cyclically exchanges angular momentum between its outer convective layers 
and the inner part. Causing cyclic spin-up and spin-down of the companion's outer layers,
this leads to an oscillation of the oblateness and the gravitational quadrupole 
moment of the companion. When the outer layers spin up, the star becomes more oblate,
the quadrupole moment increases and $\Porb$ must decrease at fixed
total orbital angular momentum. On the other hand, if  
the outer layers spin down, the quadrupole moment decreases,
and $\Porb$ increases. 

\citet{Applegate+1994} further hypothesized that tidal dissipation supplies
the energy flows driving the convection in the rotating companion. 
The resulting magnetic activity
and the mass loss due to the wind driven by pulsar irradiation contribute to a spin torque
that holds the stellar companion slightly out of synchronous rotation giving rise
to tidal dissipation and thus heats the companion internally.

\citet{Lanza+1998} proposed an extension of Applegate's model. 
Instead of the internal angular momentum redistribution considered 
by \citet{Applegate1992}, they argued that a variation of the azimuthal magnetic
field can also produce quadrupole moment changes. They 
invoked cyclic exchanges between kinetic and magnetic energy  
within the convective zone during the magnetic cycle
to cause the modified distribution of angular momentum within the star.

To better estimate the duration of the magnetic activity cycle coupling to
the observed orbital-period modulation of \psr, we compared the
timing solution to the following simple model. 
We described the change in orbital frequency $\Delta\forb(t)$
around its nominal value (according to $\Porb$ of Table~\ref{t:tim})
by a term linearly varying with time~$t$ (measured as difference from $\Tasc$ in Table~\ref{t:tim}),
and an oscillatory term with modulation period $P_{\rm mod} \equiv 1/\nu_{\rm mod} $ 
and derivative $\dot{P}_{\rm mod} \equiv - \dot{\nu}_{\rm mod} /\nu_{\rm mod}^2$,
\begin{equation}
  \Delta f_{\rm orb}(t) = \dot{f}_{{\rm orb},\star} \; t 
  + A \,\sin\left[2\pi \nu_{\rm mod} t + \pi\dot\nu_{\rm mod} t^2 + \psi \right] \,,
  \label{e:forbmodel}
\end{equation}
where $A$ and $\psi$ are the $\Porb$-modulation amplitude and phase, respectively.
Using chi-square minimization, the best-fit results we found for this modulation model along with
their formal $1\sigma$ statistical uncertainties are given in Table~\ref{t:mod}.
In particular, we estimated a modulation period $P_{\rm mod} = 4.2$\,yr and 
a modulation amplitude \mbox{$ A\Porb/(2\pi) = \Delta\Porb/\Porb = 2.3\times10^{-7}$}. 
From the inferred value for $\dot{f}_{{\rm orb},\star}$ we derived the 
residual orbital-period derivative $\dot{P}_{{\rm orb},\star}$ given in Table~\ref{t:mod}.
However, we caution that substantial systematic errors are expected,
given that the reduced chi-square value of the fit was a few times greater than unity 
and secondary minima existed.
Nonetheless, as illustrated in panel~E of Figure~\ref{f:tim}, this model 
following \Eref{e:forbmodel} allows one to qualitatively reproduce the 
observed orbital-period variation.

These results are to be compared to the GQC model, where
a change in orbital period by $\Delta \Porb$ 
is related to a change $\Delta Q$ of the companion's quadrupole moment~$Q$ 
via \citep{Applegate+1987},
\begin{equation}
  \frac{\Delta\Porb}{\Porb} = -9 \frac{\Delta Q}{M_{\rm c}\,a^2} \,.
\end{equation}
While $Q$ likely has a complex dependency on how mass, magnetic fields 
and rotational angular velocity are distributed within the star, 
it is dominated by the mass distribution in the outer layers of
the companion where the centrifugal acceleration is largest. 
To estimate the resulting change $\Delta Q$ from the transfer of angular momentum 
to the outer part of the star, \citet{Applegate1992} considered a thin
shell of mass $M_{\rm s}$ and radius $R_{\rm c}$ whose angular velocity~$\Omega$
will change by $\Delta\Omega$, giving \citep{Applegate+1994},
\begin{equation}
  \frac{M_{\rm s}}{M_{\rm c}} \frac{\Delta\Omega}{\Omega}
   = \frac{G M_{\rm c}}{2 R_{\rm c}^3} \; \frac{a^2}{R^2_{\rm c}}\: \frac{\Porb^2}{4\pi^2} \;\frac{\Delta\Porb}{\Porb} \,.
\end{equation}
Assuming corotation of the tidally locked companion, we obtain for the \psr{} system,
\begin{equation}
	 \frac{\Delta\Omega}{\Omega}  = 9.7\times10^{-7}\;  \frac{M_{\rm c}}{M_{\rm s}} \,.
\end{equation}
\citet{Applegate1992} found observational results to be typically fit well for a shell 
of mass $M_{\rm s} \approx 0.1 M_{\rm c}$.
Adopting this value here implies $\Delta\Omega/\Omega \approx 10^{-5}$.
In Applegate's model, the magnetic field generation and angular velocity
variation operates cyclically with period~$\Pmod$ (the cycle of
the orbital-period modulation).
While no further details of this activity cycle were specified,
\citet{Applegate1992,Applegate+1994} assumed that the energy needed for the transfer
of angular momentum would require an associated change in the star's luminosity of 
\begin{equation}
  \Delta L = \frac{\pi}{3} \frac{G\; M_{\rm c}^2}{\Pmod} \;\frac{a^2}{R_{\rm c}^3} \frac{\Delta\Omega}{\Omega} \;\frac{\Delta\Porb}{\Porb} \,.
  \label{e:DeltaL}
\end{equation}
Using the above estimated value for $\Pmod$, the associated luminosity change
is $\Delta L \approx 4.1\times10^{28}$\,erg\,s$^{-1}$. This is only about $0.03$\% of the intrinsic
luminosity of the companion, which we estimated in Section~\ref{ss:energetics}. 
Therefore, the companion should be easily capable of providing the 
required energy to power the orbital-period variations.
Even if the true component masses were slightly different from those assumed, this 
conclusion remains robust. This is illustrated in Figure~\ref{f:massmass}, where 
the required fractional luminosity is shown to vary by less than an order of magnitude
over the permitted region in the mass-mass diagram.

\begin{deluxetable}{ll}
\tablewidth{\columnwidth}
\tablecaption{\label{t:mod} Estimated Orbital-period Modulation Model Parameters}
\tablecolumns{2}
\tablehead{
\colhead{Parameter} &
\colhead{Value}
}
\startdata
Modulation amplitude, $A$ (s$^{-1}$) \dotfill & $8.48(2)\times 10^{-11}$\\
Modulation period, $P_{\rm mod}$ (yr) \dotfill & $4.15(1)$\\
Modulation period derivative, $\dot{P}_{\rm mod}$ (s s$^{-1}$) \dotfill  & $-0.87(1)$\\
Modulation phase, $\psi$ (rad) \dotfill & $5.64(1)$\\
Residual orbital-period derivative, $\dot{P}_{{\rm orb},\star}$ (s s$^{-1}$) \dotfill & $-1.57(2)\times10^{-11}$ \\
\enddata
\tablecomments{Numbers in parentheses are only statistical, formal $1\sigma$ uncertainties 
in the last digits.}
\end{deluxetable}

For other classes of close binaries, such as the RS Canum Venaticorum systems, 
\citet{Lanza2005} \citep[and ][]{Lanza2006} found a discrepancy of Applegate's 
GCQ model: The required surface angular velocity variations 
led to an energy dissipation rate in the turbulent convection zone of the
secondary star exceeding the stellar luminosity. For \psr{}, with lower 
$\Delta\Porb/\Porb$ and $\Delta\Omega/\Omega$, this is not the case. 
Applying Equation~(26) of \citet{Lanza2006}, we infer a ratio between the 
dissipated power and the stellar luminosity of only a few percent, further supporting Applegate's 
GCQ mechanism. However, if the true companion radius was much smaller than the one we assumed above,
this energy balance could eventually become questionable.
But this would also imply that the companion was actually not near filling its Roche lobe,
which is currently not supported observationally \citep{Romani+2011-J2339}.

Moreover, on the timescale of the observational data, 
the component masses and the orbital angular momentum 
are approximately constant as presumed by the GQC model. This implies that
the observed orbital-period variation must also lead to a change in the
orbital separation: $\dot{a}/a = 2 \dot{P}_{\rm orb}/\Porb$. Taking the largest
$\dot{P}_{\rm orb}$ we observed, this would  give \mbox{$\dot x \simeq 2\times10^{-14}$\,lt-s s$^{-1}$}. 
This is still smaller than the measured upper limit on $\dot x$ provided in Table~\ref{t:tim} and
thus this aspect of the model is also in line with our timing measurements.

The above analysis of the orbital-period evolution (Table~\ref{t:mod}) indicates that aside 
from the modulation, a long-term change of $\dot{P}_{{\rm orb},\star} \approx 10^{-11}$\,s s$^{-1}$
might still remain. If this residual orbital-period derivative was mostly due to mass loss,
the next largest contribution as estimated in Section~\ref{ss:Porbmod}, 
then the rate would be 
\mbox{$\dot{M}_{\rm c}\approx 2\times10^{-8}\,M_\odot$yr$^{-1}$}.
To match this mass-loss rate the estimates made
in Section~\ref{ss:Porbmod} would require only little modification
of the assumed irradiation efficiency or a slight deviation from the pulsar's presumed isotropic emission. 

Overall, we conclude that the GQC theory offers a compatible explanation of the 
orbital-period variation observed for \psr{}.

\section{Conclusions}\label{s:conclusions}

Using the available \Fermi-LAT data spanning more than six years, 
we carried out a precision gamma-ray timing analysis of the redback-type 
pulsar binary \psr{}. Most notably, the results revealed a long-term
modulation of the $4.6$\,hr binary orbital period.
We found that this observed phenomenon can be explained 
by variations of the gravitational quadrupole moment of the companion,
through a mechnism proposed by \citet{Applegate1992}. 

Since Applegate's model offers a compatible explanation, it implicates that 
the companion star must have a sizeable outer convective zone, where cyclic magnetic activity causes
the quadrupole moment changes leading to the observed orbital-period modulation.
Additional evidence for such convective fluctuations is provided by
our measured residual orbital eccentricity, found to be
higher than theoretically predicted for such a tight binary \citep{Phinney1992}.
The strong irradiation and possibly also the tidal interaction with the close-by pulsar 
may drastically affect the internal structure of the companion making it similar to 
that of a main-sequence low-mass star with a convective envelope.
While its optical spectrum is in fact consistent with 
a late-type star \citep{Romani+2011-J2339}, the companion however appears
to be less dense than a main-sequence star of the same mass.

Combined with the pulsar ephemeris, future optical observations of the 
companion can help to more tightly constrain the system parameters,
e.g., their component masses. As another interesting prospect, 
such measurements might also be able to provide evidence
for a variation in accordance with stellar activity.
This might also allow further tests of the GQC theory, given the high timing precision with which 
we can measure the orbital-period variation. 
Ultimately, one might hope to identify the type of stellar magnetic dynamo in action
or the physical origin of the differential rotation within the companion's convective layers,
currently observationally difficult to access otherwise.

Surprisingly many black widow and redback pulsar binaries have been discovered 
in targeted radio searches of unidentified LAT sources.
However, the fact that many of those systems also show significant radio eclipses and 
variations in their orbital parameters often makes it still extremely difficult to also detect 
the gamma-ray pulsations. That is because the short-term radio timing solutions 
cannot be extended immediately backwards to cover the full LAT data time span
since \Fermi{}'s launch. 
As demonstrated for \psr{}, the method presented here can help address 
this problem. Besides revealing the gamma-ray pulsations,
by making full use of the great potential of the exquisite LAT data,
this also opens the unique possibility for long-term monitoring and insights 
into the complex interplay of effects influencing the orbital evolution
of such irradiating pulsar binary systems.

\acknowledgements

We thank fellow members of the LAT collaboration for helpful discussions and 
Lucas Guillemot, Massimiliano Razzano, Paul Ray, and Roger Romani for helpful comments
on the manuscript. This work was supported by the Max-Planck-Gesellschaft~(MPG), 
as well as by the Deutsche Forschungsgemeinschaft~(DFG) through 
an Emmy Noether Research Grant, No. PL~710/1-1 (PI: Holger~J.~Pletsch). 

The \textit{Fermi} LAT Collaboration acknowledges generous ongoing support
from a number of agencies and institutes that have supported both the
development and the operation of the LAT as well as scientific data analysis.
These include the National Aeronautics and Space Administration and the
Department of Energy in the United States, the Commissariat \`a l'Energie Atomique
and the Centre National de la Recherche Scientifique/Institut National de Physique
Nucl\'eaire et de Physique des Particules in France, the Agenzia Spaziale Italiana
and the Istituto Nazionale di Fisica Nucleare in Italy, the Ministry of Education,
Culture, Sports, Science and Technology (MEXT), High Energy Accelerator Research
Organization (KEK) and Japan Aerospace Exploration Agency (JAXA) in Japan, and
the K.~A.~Wallenberg Foundation, the Swedish Research Council and the
Swedish National Space Board in Sweden.
 
Additional support for science analysis during the operations phase is gratefully acknowledged from the Istituto Nazionale di Astrofisica in Italy and the Centre National d'\'Etudes Spatiales in France.

\bibliographystyle{apj}

\bibliography{orbvarJ2339} 

\end{document}